\documentclass[a4paper]{article}

\usepackage{multirow}
\usepackage{hyperref}
\usepackage{balance} %
\usepackage{INTERSPEECH2021}
\graphicspath{images/}

\title{Distance-Based Sound Separation}
\name{Katharine Patterson, Kevin Wilson, Scott Wisdom, John R. Hershey}
\address{
  Google Research, Cambridge MA}
\email{\{kappi, kwwilson, scottwisdom, johnhershey\}@google.com}

\begin{document}

\maketitle
\begin{abstract}
We propose the novel task of 
\emph{distance-based sound separation}, where sounds are separated based only on their distance from a single microphone. 
In the context of assisted listening devices, proximity provides a simple criterion for sound selection in noisy environments that would allow the user to focus on sounds relevant to a local conversation.  We demonstrate the feasibility of this approach by training a neural network to separate near sounds from far sounds in single channel synthetic reverberant mixtures, relative to a threshold distance defining the boundary between near and far.  With a single nearby speaker and four distant speakers, the model improves scale-invariant signal to noise ratio by 4.4 dB for near sounds and 6.8 dB for far sounds.
\end{abstract}
\noindent\textbf{Index Terms}: distance-based sound separation

\section{Introduction}

Extracting estimates of clean speech in the presence of interference is a long-standing research problem in signal processing. This task is referred to as \textit{speech enhancement} when the interference is non-speech, and \textit{speech separation} when the interference is speech. More generally, \emph{sound separation} refers to the extraction of a subset of sounds from a mixture of sounds.   

Hearing aids and other assisted listening devices are an important application for these methods, e.g.\ \cite{wang2008time, wang2019low, fedorov2020tinylstms}, with a typical task being to aid in conversations held in a noisy space. 
But implementing sound separation for assisted listening is challenging, requiring low algorithmic latency and limited computation and memory. Further, it is not always clear which sounds the user would like to hear.  

The problem of selecting which sounds to enhance has been approached by focusing only on speech \cite{wang2019low}, or otherwise classifying the sounds \cite{kong2020source, wisdom2021sparse}, as well as using visual input \cite{hussain2017towards, ephrat2018looking, gogate2020cochleanet} to select the sounds of interest.
However, these methods aren't always appropriate.  Selective listening methods may be cumbersome and require user effort to control which sounds are enhanced. 
They also typically exclude sounds other than speech that the user may want to hear, such as nearby music, the clinking of wine glasses at dinner, or the tell-tale sound of a dropped set of keys.  Class-based methods that try to include non-speech sounds can fail whenever there are interfering sounds of the same classes as the desired sounds. 

\begin{figure}
    \centering
    \includegraphics[width=\linewidth]{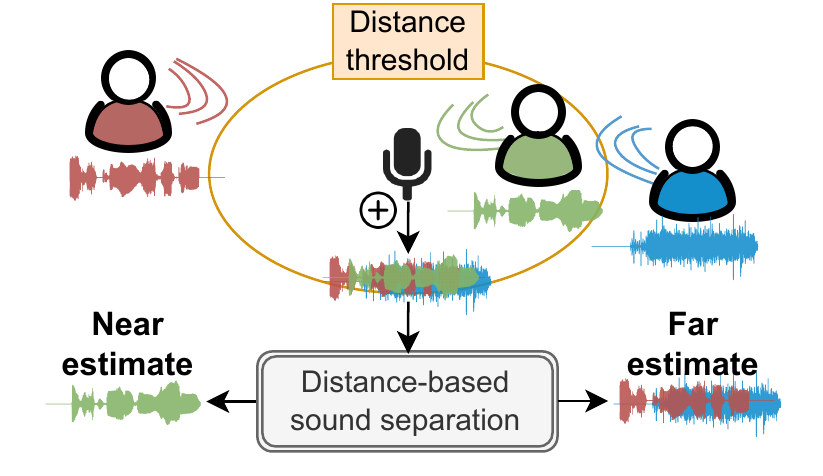}
    \caption{Illustration of the proposed task of distance-based sound separation.}
    \label{fig:illustration}
\end{figure}

We propose an alternative approach called \emph{distance-based sound separation}, illustrated in Figure \ref{fig:illustration}, in which we assume the user would like to hear any sounds that occur within a local region around them, and block sounds coming from farther away.  A system that accomplishes this would allow the user to engage in normal conversations without the interference of a crowded environment, and without becoming deaf to non-speech sounds in their immediate area.  One hypothesis is that there are cues for distance that are related to superficial characteristics of the sound, rather than to its fine spectro-temporal structure.  If so, a system that relies on such cues might have both a computational advantage, and an advantage in terms of generalization, compared to a system that has to perform deep pattern recognition in order to separate sounds.  Although humans have the perceptual ability to estimate the relative distance of sounds \cite{kolarik2016auditory}, there has been no prior work showing that distance-based separation is feasible.

Possible cues for distance perception can be traced to physical effects. %
The intensity $I$ of the \emph{direct path} component of the sound varies with distance $d$ according to inverse-square law $I \propto 1/d^2$.  In contrast, in an enclosed area, the intensity of the sound's reverberation, is roughly independent of its distance to the microphone.  Thus the \emph{direct-to-reverberation} (DRR) ratio decreases with distance, and has been shown to be a cue for human distance perception \cite{kolarik2016auditory}.  There is also a \emph{proximity effect} with directional microphones in which low-frequencies are more accentuated for closer sources, but the effect is only strong for sources less than one meter away \cite{milanova2001proximity,milanov2000proximity}. Absorption by the air is a frequency-dependent effect of distance, and plays a strong role over distances beyond tens of meters \cite{harris1966absorption}.
Other effects of distance come from spatial effects that could be detected using an array of microphones.  In the present work, however, we focus on what can be inferred using a single microphone.  

To validate the concept of distance-based sound separation, we train neural network separation models using mixtures of near and far sounds, where the acoustic properties of the sounds have been emulated using an acoustic room simulator.  This allows us to have ground-truth targets for the near and far signals relative to a distance threshold.  In this work, we focus on the case where all sounds, both near and far, are speech.  We leave the case where there are non-speech sounds for future work.  Our results show that it is possible to perform separation solely based on distance.  In particular, in some scenarios, such as with a distance threshold of 1.5 m, a single nearby speaker, and as many as four distant speakers, distance-based sound separation can achieve improvements of up to 4.4 dB in scale-invariant signal to noise ratio.

\section{Related Work}

Beamforming \cite{van2004optimum} uses multiple microphones and generally uses the direction of arrival (DOA) of sources as a cue to separate them. Near-field beamforming methods can additionally estimate source distance using the spherical nature of acoustic wavefronts \cite{ryan1997near, zheng2004robust}, though they can only distinguish distances of nearby sounds within a range limited by the size of the microphone array.
In contrast, our method can learn to operate over a wider range of distances.
Distance estimation has also been investigated using neural networks with two microphones \cite{yiwere2017distance,ferguson2018sound, yiwere2019sound,krause2021joint}. However, our method only requires a single microphone, and it performs separation.

Prior work has demonstrated the feasibility of estimating reverberation parameters such as T60 and DRR \cite{kolarik2016auditory}. Neural networks can also be trained to estimate reverberation parameters from a single microphone \cite{gamper2018blind, bryan2020impulse}, as well as room volume \cite{genovese2019blind}. The success of these methods suggests that neural networks are able to perceive cues from raw audio to accurately predict properties of acoustic transfer functions, and it is likely that neural networks could also be trained to estimate the distance of a source. Rather than predict reverb parameters, our proposed model implicitly leverages acoustic cues to identify near and far sources in order to separate them.

Single-channel dereverberation methods \cite{ nakatani2006harmonicity, habets2010speech, luo2018real, ernst2018speech} solve a somewhat related task to distance-based sound separation, in that they try to separate the direct path of a reverberant signal (the shortest path of sound propagation) from the reverberant components. In contrast, the goal of distance-based sound separation is to preserve the reverberance of sources, and to group sources together based only on whether they are near or far based on a distance threshold. Also, dereverberation methods often make the assumption of a single reverberant source, while our method can handle multiple sources.

To our knowledge, the work introduced here is the first to propose and demonstrate the separation of speech from a single microphone based solely on distance cues.

\section{Methods}
\subsection{Model}
To separate based on distance, we experimented with an architecture  
\cite{differentiable_consistency}, that uses short-time Fourier transform (STFT) masking. For the STFT, we use a 32 ms square-root Hann window with 16 ms hop. The 0.3-power-compressed magnitude of the STFT, $Y$, of the time-domain mixture, $y$, is fed as input to $L$ layers of uni-directional LSTMs \cite{hochreiter1997lstm} with $N$ units each. Then, a fully-connected layer with a sigmoid activation is applied to create two masks for the input STFT $Y$, $M_\mathrm{near}$ and $M_\mathrm{far}$. To ensure STFT consistency \cite{differentiable_consistency}, the masked STFTs for near and far are each passed through inverse and forward STFT operations: $\hat{X}_\mathrm{near| far}=\mathrm{STFT}\{\mathrm{iSTFT}(M_\mathrm{near|far} \odot Y)\}$.

The training loss is mean-squared error between 0.3-power-compressed magnitude of target $X$ and estimate $\hat{X}$ \cite{differentiable_consistency}:
\begin{equation}
  \label{eq:loss}
  \mathcal{L}(X, \hat{X})  = \sum_{f,t} 
  \big(
    |X_{f,t}|^{0.3} - |\hat{X}_{f,t}|^{0.3}
  \big)^2
\end{equation}
We use weights of 0.8 on the near loss and 0.2 on the far loss,
$0.8\,\mathcal{L}(X_\mathrm{near}, \hat{X}_\mathrm{near})
+
0.2\,\mathcal{L}(X_\mathrm{far}, \hat{X}_\mathrm{far}),$
which encourages the model to focus on the performance of near targets, which is more likely to be desired as the output in a practical application.
For all experiments, we used the Adam optimizer \cite{kingma2014adam} with a learning rate of $3 \times 10^{-5}$, batch size 128, and trained for one million steps on 16 Google Cloud TPU v3 cores.

\subsection{Acoustic Simulation}

To create a large amount of training data, we use an image-method \cite{simulating_acoustics} acoustic room simulator with frequency-dependent wall filters \cite{wang2021sequential} to generate reverb impulse responses (RIRs) for rooms with varied acoustic properties, with randomized microphone and source locations.  Basic distance-related phenomena such as the amplitude and DRR effects are well simulated. 
However, our simulation does not model 
the proximity effect and the air absorption effect, which might provide additional cues for distance-based separation in the real world.
Clean speech recordings are randomly assigned to the source locations within each room and convolved with the corresponding RIRs. 
We created training examples for each room by combining all sources within a threshold distance into a near target $x_\mathrm{near}$, and all sources beyond a threshold distance into a far target $x_\mathrm{far}$. The near and far targets are added to create mixture $y=x_\mathrm{near}+x_\mathrm{far}$, and fed into the model to be separated.

\begin{figure}
    \centering
    \begin{minipage}{0.58\linewidth}
        \centering
        \includegraphics[width=\linewidth]{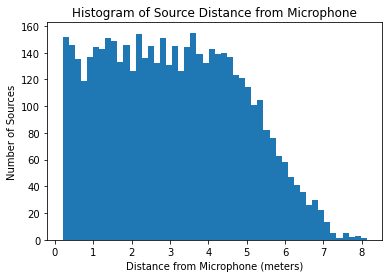}
    \end{minipage}\hfill
    \begin{minipage}{0.42\linewidth}
        \centering
        \includegraphics[width=\linewidth]{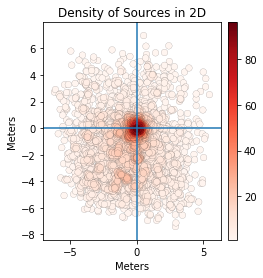}
    \end{minipage}
    \caption{Source distribution by distance from microphone, and spatial distribution in 2D.}
    \label{fig:source_distribution}
\end{figure}

Within each of the randomly generated rooms, we randomly generate 5 speaker locations and 1 microphone location, such that the distance between the microphone and each source was uniformly distributed, subject to the rejection of samples falling outside each room. The resulting distribution of distances from the microphones is illustrated in Figure \ref{fig:source_distribution}. Note that the spatial distribution differs from what might be expected in, for example, a restaurant setting, where the average number of sources at a given distance radius will increase as the distance increases.

\section{Experiments}

\subsection{Data Preparation}
\label{ssec:data_prep}

For the experiments, we use speech recordings from the Libri-light dataset \cite{librilight} for training, with validation, and test partitions coming from LibriSpeech \cite{panayotov2015librispeech}, so that speaker IDs are unique for each partition.  
For our synthetic room data, rooms are generated with dimensions varying from 3.0$\times$4.0$\times$2.13 meters to 7.0$\times$8.0$\times$3.05 meters. 

During training, randomly chosen Libri-light clips are reverberated and mixed according to a randomly chosen RIR to create clips that are 10 seconds in duration.
Source utterances shorter than 10 seconds are offset at random intervals within the 10 second clip, and source utterances longer than 10 seconds are clipped to a random 10 second interior segment.

The source and microphone locations are used to determine which sources are near or far in relation to the microphone in the given room, for a given threshold distance. To vary the number of sources, we apply a \textit{source presence probability} (SPP) to each source, so that the total number of sources in a room varies from 0 to 5, with a distribution dependent on the chosen SPP.

Because the rooms are generated independently of a specific distance threshold, the choice of distance threshold affects the number of sources considered near versus far for any given room.
In particular, it affects the fraction of examples where all the sources are considered near \textit{or} all sources are considered far, leading to silent far targets \textit{or} silent near targets. 
Distances thresholds of 0.8/1.5/3.0 meters result in training data with silent far target fractions of 0.0/0.001/0.04, respectively and silent near target fractions of 0.59/0.30/0.03 respectively.
It is notable that some thresholds result in a majority of training examples containing a silent target, even for an SPP of 1.0.

\subsection{Metrics}

For training examples where both near and far targets have sources present (i.e.\ neither target is silent), we use scale-invariant SDR improvement \cite{LeRoux2018a} (SI-SDRi), $\text{SI-SDRi}(x, \hat{x})=\text{SI-SDR}(x, \hat{x}) - \text{SI-SDR}(x, y)$. SI-SDR measures signal fidelity with respect to a reference signal while allowing for a gain mismatch:
\begin{equation}
\text{SI-SDR}(x, \hat{x}) = 10 \log_{10}
\left(
    {\|\alpha x\|^{2}}
    /
    {\|\alpha x - \hat{x}^{s}\|^2}
\right),
\end{equation}
where $\alpha$ = argmin$_a \| a x - \hat{x} \|^{2} = x^{T} \hat{x} / \|x\|^{2}$.

SI-SDRi diverges when one of the targets is silent, because the non-silent target will be exactly equal to the input mixture and achieve $\infty$ dB SI-SDR, and the silent target will have $-\infty$ dB SI-SDR. This makes any improvement calculation meaningless.
For examples with a silent near target, we use a noise reduction metric, to measure how much of the far sound leaks into the silent near output:
\begin{equation}
\label{eq:noise_reduction}
\text{NoiseReduction}(y, \hat{x}_\mathrm{near}) = 10 \log_{10}
\left(
{\|y\|^{2}}
/
{\|\hat{x}_\mathrm{near}\|^2}
\right),
\end{equation}
which measures the power reduction of the separated output $\hat{x}_\mathrm{near}$ relative to the input audio mixture power $y$.

\section{Results} %
\label{sec:results}

Results are shown in Table~\ref{tab:results} for an evaluation set of 1000 examples with a distance threshold of 1.5 meters, model size of $L=4$ and $N=400$ (6.3M parameters), SPP of 0.5, and with all 5 speakers present. They are bucketed according to number of near speakers (1 near speaker implies 4 far speakers, etc).  

We compare our distance-based separation network to a conventional speech separation baseline in which we train the same network architecture (but with more masking outputs and a permutation-invariant loss) to output 5 separate source estimates, one for each possible speaker.  To upper-bound the performance of this baseline, we use oracle knowledge of the true near and far targets to find the best grouping of the 5 separate source estimates into near and far estimates.  In all but one condition, direct distance-based separation outperforms this upper bound on our ``separate-all'' baseline.

\begin{figure*}
    \centering
    \begin{minipage}{0.24\linewidth}
        \centering
         \includegraphics[
         height=3.5cm,
         ]{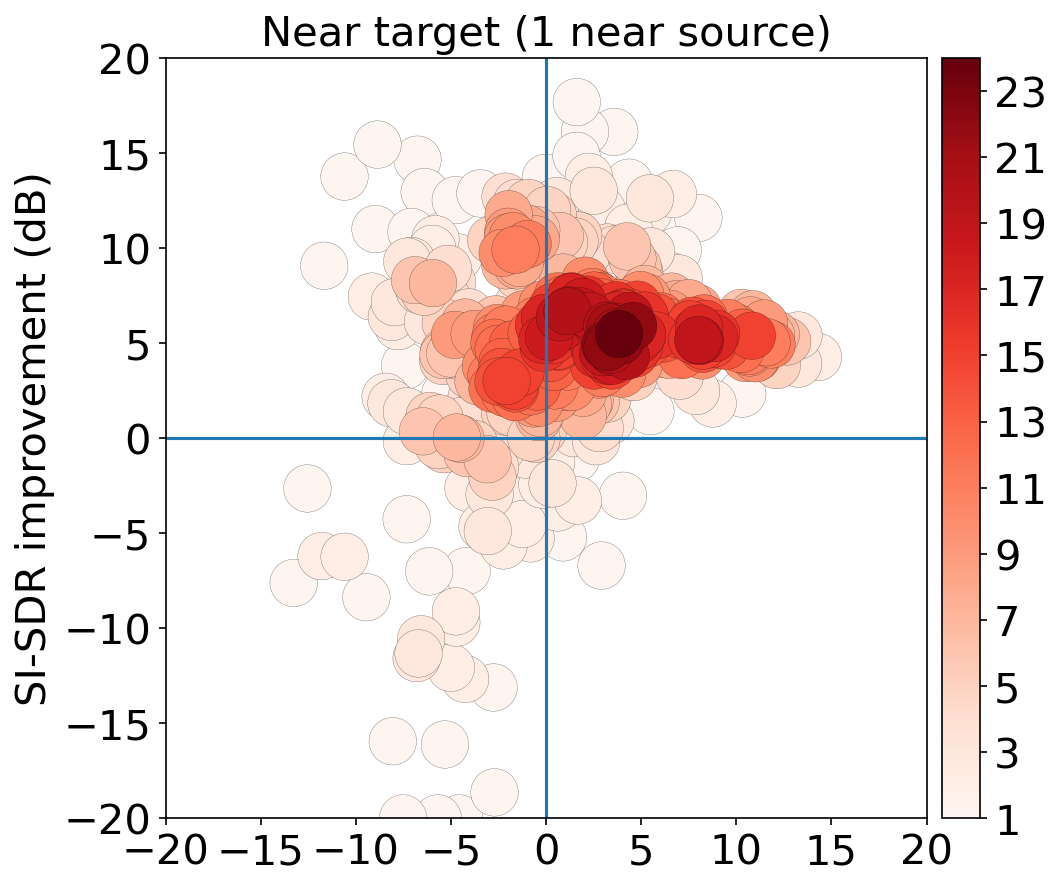}
    \end{minipage}\hfill
    \begin{minipage}{0.24\linewidth}
        \centering
        \includegraphics[
        height=3.5cm,
        ]{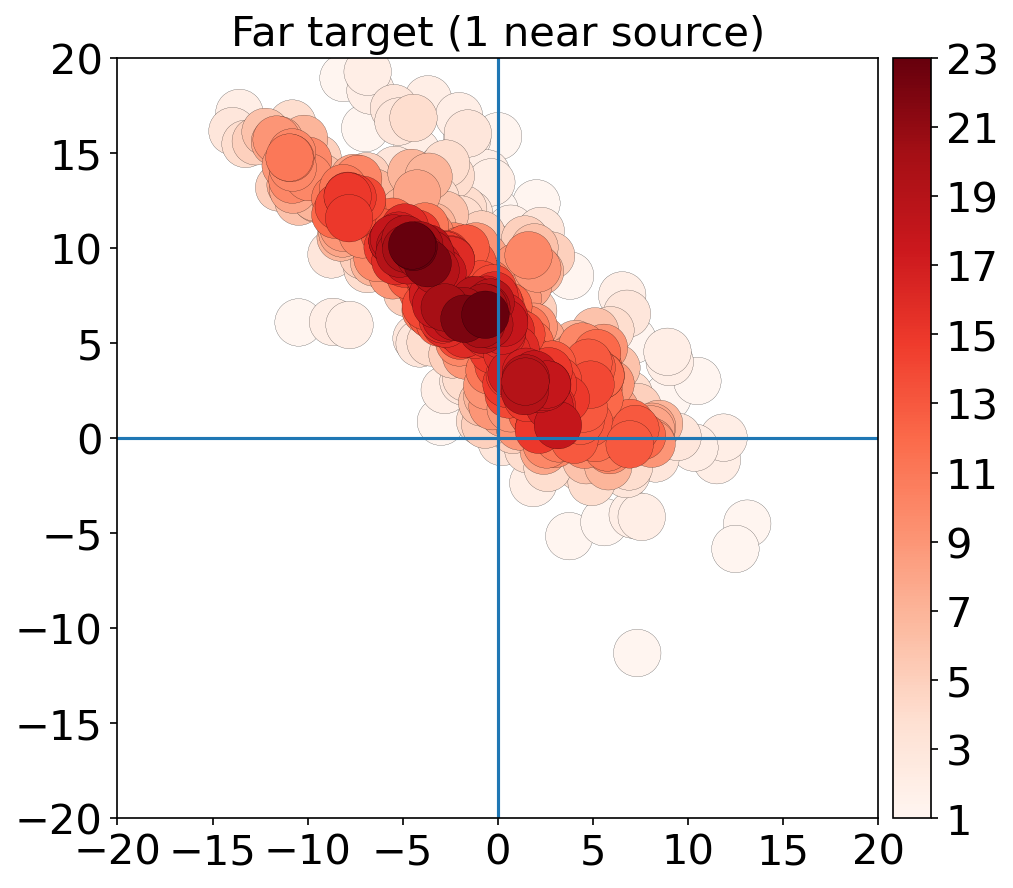}
    \end{minipage}
    \vline
    \begin{minipage}{0.24\linewidth}
        \centering
        \includegraphics[
        height=3.5cm,
        ]{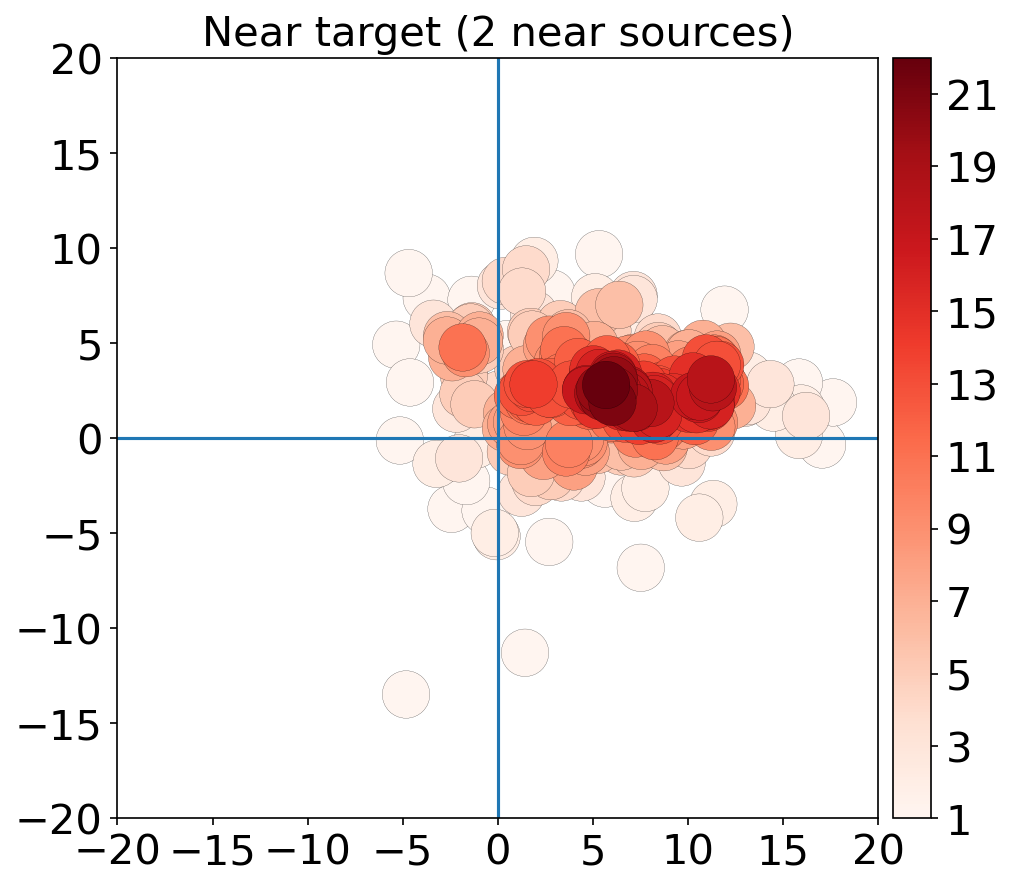}
    \end{minipage}\hfill
    \begin{minipage}{0.24\linewidth}
        \centering
        \includegraphics[
        height=3.5cm,
        ]{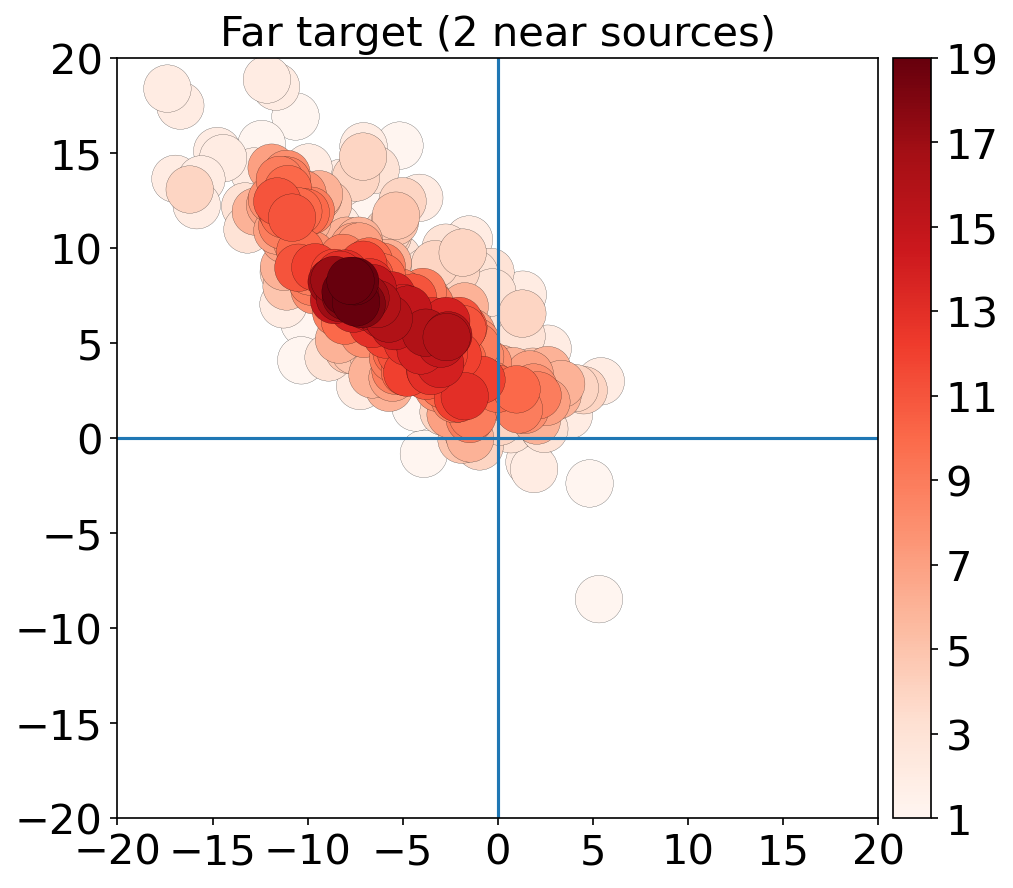}
    \end{minipage}
    \hrule
    \begin{minipage}{0.24\linewidth}
        \centering
        \includegraphics[
        height=3.725cm,
        ]{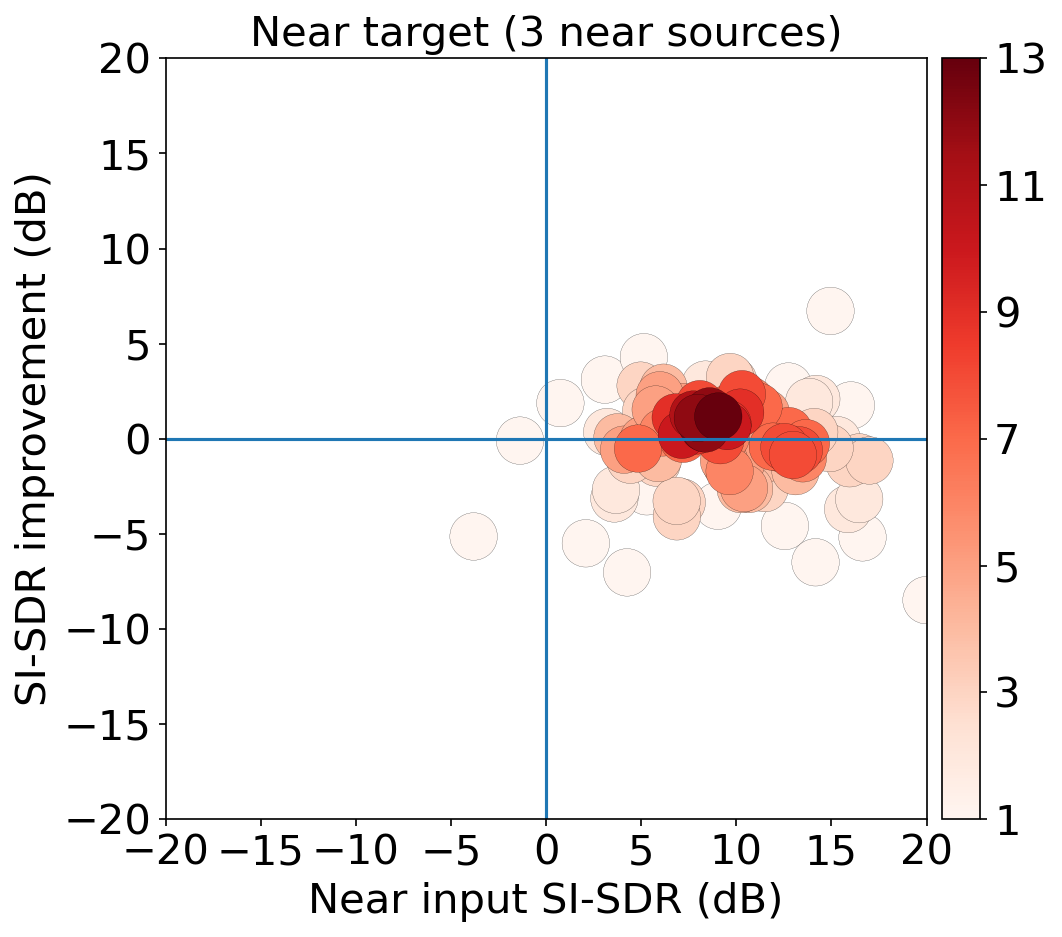}
    \end{minipage}\hfill
    \begin{minipage}{0.24\linewidth}
        \centering
        \includegraphics[
        height=3.725cm,
        ]{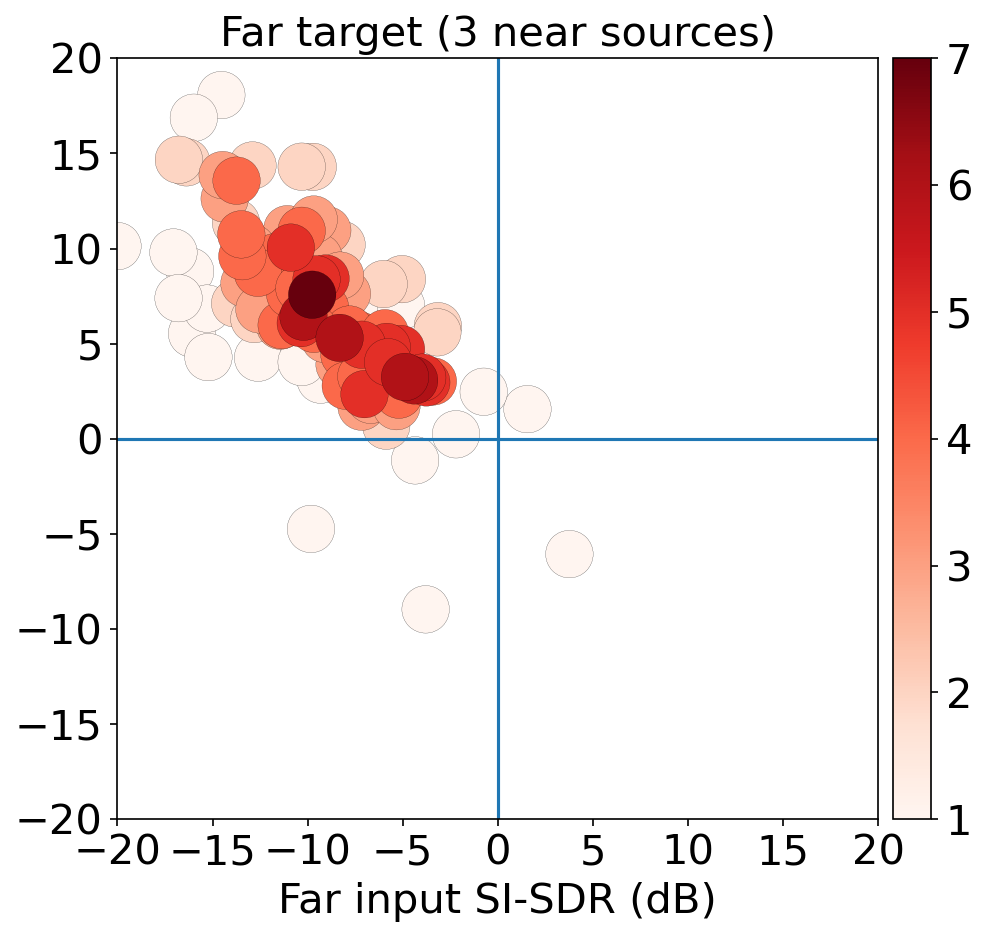}
    \end{minipage}
    \vline
    \begin{minipage}{0.24\linewidth}
        \centering
        \includegraphics[
        height=3.725cm,
        ]{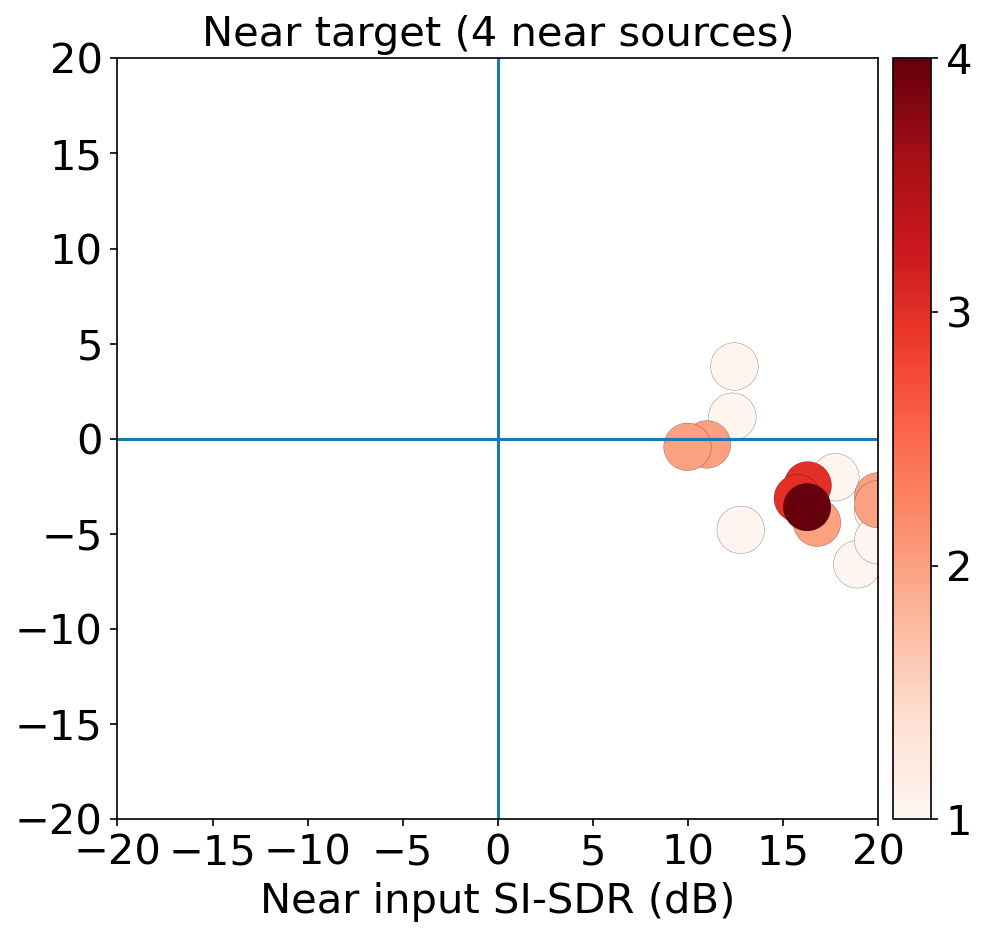}
    \end{minipage}\hfill
    \begin{minipage}{0.24\linewidth}
        \centering
        \includegraphics[
        height=3.725cm,
        ]{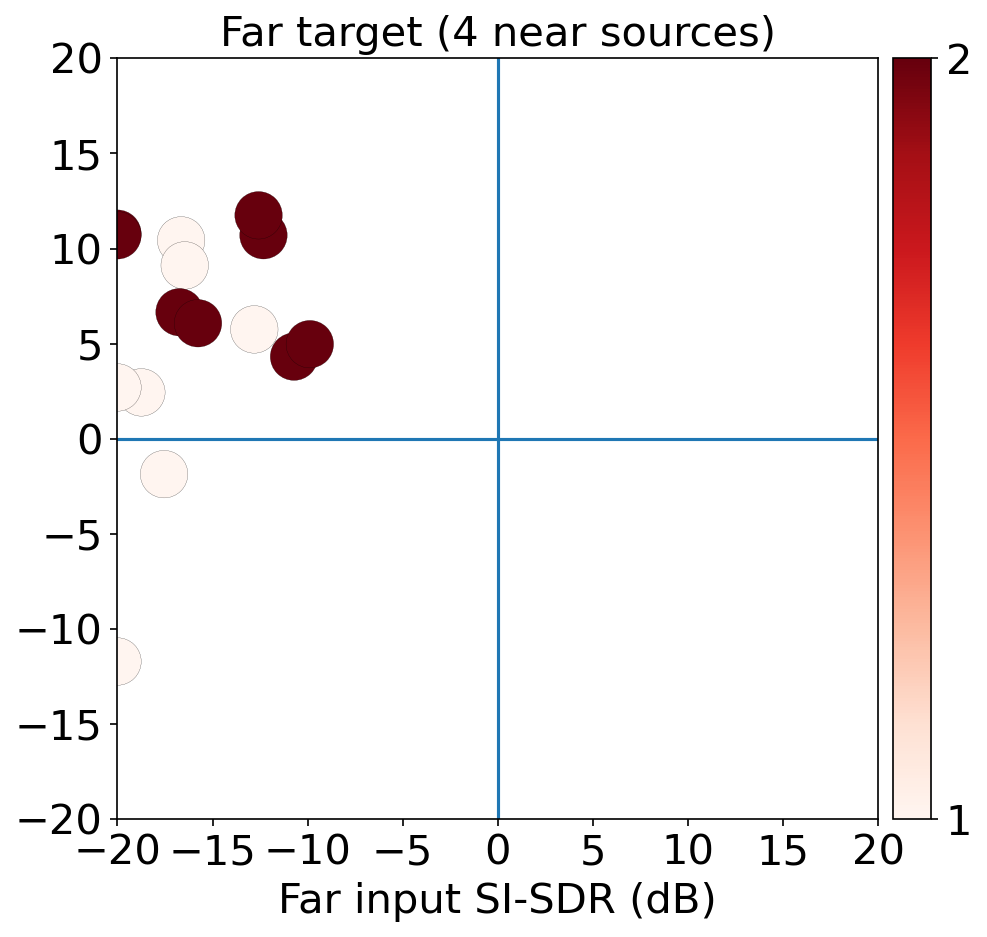}
    \end{minipage}
    \caption{SI-SDRi in dB as a function of input SI-SDR for near and far targets, for the model evaluated in Table \ref{tab:results} on test examples with a distance threshold of 1.5 m. Results are divided by the number of sources in the near target.}
    \label{fig:1_near_results}
\end{figure*}

Figure \ref{fig:1_near_results} shows scatter plots of near input SI-SDR versus SI-SDR improvement for individual examples.
The mean noise reduction (\ref{eq:noise_reduction}) for this model across the 271 evaluation examples with silent near targets (n/a in Table \ref{tab:results}) is 48.9 dB.

Demos are online at \url{https://google-research.github.io/sound-separation/papers/distance-based-separation-interspeech2022}.

\begin{table}[th]
  \caption{Mean SI-SDRi for distance threshold of 1.5~m, LSTM depth $L=4$, width $N=400$, and training SPP of 0.5, on an eval set of 1000 5-speaker examples with SPP of 1.0.  "Distance-based" is our method of directly training to separate near sounds from far.  ``Separate-all'' is a baseline in which same network architecture is trained to output 5 individual sources, which are then grouped into near and far estimates.
  }
  \label{tab:results}
  \centering
\begin{tabular}{c|cc|cc|l}
 \multicolumn{1}{c|}{\multirow{2}{1cm}{\textbf{\# near sources}}} & \multicolumn{2}{c|}{\textbf{Distance-based}} & \multicolumn{2}{c|}{\textbf{Separate-all}} \\
 & \textbf{Near} & \textbf{Far} & \textbf{Near} & \textbf{Far} & \textbf{(\#)} \\ \hline
1 & {\bf 4.4} & {\bf 6.8} & 2.5 & 3.9 & (372) \\
2 & {\bf 2.2} & {\bf 7.1} & 0.4 & 4.7 & (248) \\
3 & {\bf -0.3} & {\bf 6.7} & -2.1 & 4.9 & (92) \\
4 & {\bf -2.6} & 5.5 & -6.6 & {\bf 5.7}  & (15) \\
\end{tabular}
\end{table}

\subsection{Effect of distance threshold}

The results of training and evaluating with different distance thresholds is shown in Table \ref{tab:distance_effects}.

\begin{table}[th]
  \caption{Performance as a function of distance threshold, for models identical to that in Table \ref{tab:results}, but trained with varied distance thresholds. SI-SDRi is given for all examples with 1 non-silent near source. Noise reduction is calculated only for the 271 examples with silent near targets.}
  \label{tab:distance_effects}
  \centering
\begin{tabular}{c|cc|c}
\multirow{2}{*}{\textbf{Distance (m)}} & \multicolumn{2}{c|}{\textbf{SI-SDRi, 1 near source}} & \multirow{2}{*}{\textbf{\begin{tabular}[c]{@{}c@{}}Noise \\ reduction\end{tabular}}} \\
 & \textbf{Near} & \textbf{Far} &  \\ \hline
0.8 & 4.0 & 8.3 & 63.9 \\
1.5 & 4.4 & 6.8 & 48.9 \\
3.0 & 1.9 & 2.7 & 22.8
\end{tabular}
\end{table}

Performance generally degrades with increased distance threshold values, despite the fact that higher thresholds resulted in more non-silent training targets. This can be seen in Figure~\ref{fig:source_distribution} and Section \ref{ssec:data_prep}
where a threshold at 3 meters is closer to the center of the distribution than a threshold of 0.8 or 1.0, and thus will change the training examples to have fewer silent targets.

The correlation between increased performance and decreased silent targets seen in Table~\ref{tab:silent_targets_results} suggests that the degradation seen in Table~\ref{tab:distance_effects} with increased distance threshold is a true artifact of the distance, as opposed to an effect of the change in training data between models.

\subsection{Effect of model size}
All combinations of LSTM depth $L\in\{2, 4, 6\}$ and width $N\in\{200, 400, 600\}$ are evaluated, except $L=6$ and $N=600$, which was too large for our resource constraints.
Overall, larger models performed better, with LSTM width $N$ contributing more to performance increases. The SI-SDR difference between the smallest model and the largest model for distance thresholds of 0.8 m, 1.5 m, and 3.0 m was 0.9 dB near / 1.2 dB far, 1.5 dB near / 1.6 dB far, and 1.8 dB near / 1.4 dB far, respectively. Note that for increasing model size, SI-SDRi improves more for higher distance thresholds.

\subsection{Effect of varied speaker count in evaluation set}

To explore whether the model uses the relationship between the number of near sources and the number of far sources, we vary the SPP.
For instance, an SPP of 1.0 indicates that all 5 sources are present in the mixture, and in general a SPP of $p$ induces a binomial distribution $\mathcal{B}(5, p)$ over number of sources.
This also modifies the distribution of near and far speaker counts. Our baseline model evaluated in Table \ref{tab:results}
was trained with a SPP of 0.5, but 
our evaluation set used in other sections uses a SPP of 1.0 for ease of slicing data by number of near speakers. This means that training data and the evaluation data differ in terms of near and far speaker counts.

In order to study this mismatch, we also created a version of the evaluation set with a SPP of 0.5.
This provides a set with 527 silent near target examples, 411 examples with sound in near and far (i.e.\ 1-4 speakers in each target, sum $\leq$ 5), and 62 examples with silent far targets.
We also trained our baseline model with SPP of 1.0 instead of 0.5. 
Results for the baseline model and eval data at SPPs of 0.5 and 1.0 are in Table \ref{tab:speaker_count_effects}.

\begin{table}[th]
  \caption{Performance as a function of SPP, for same model as Table \ref{tab:results}, except for SPP. For Eval SPP=1, SI-SDRi is given for all 727 examples with non-silent near and far targets. Noise reduction is calculated only for the 271 examples with silent near targets. For Eval SPP=0.5, SI-SDRi is for 394 non-silent near and far target examples, noise reduction for 538 silent near targets examples. }
  \label{tab:speaker_count_effects}
  \centering
\begin{tabular}{cc|cc|c}
\multirow{2}{*}{\textbf{\begin{tabular}[c]{@{}c@{}}Model \\ SPP\end{tabular}}} & \multirow{2}{*}{\textbf{\begin{tabular}[c]{@{}c@{}}Eval \\ SPP\end{tabular}}} & \multicolumn{2}{c|}{\textbf{\begin{tabular}[c]{@{}c@{}}SI-SDRi all non-silent\end{tabular}}} & \multirow{2}{*}{\textbf{\begin{tabular}[c]{@{}c@{}}Noise \\ reduction\end{tabular}}}  \\
& & \textbf{Near} & \textbf{Far} & \\ \hline
0.5 & 1 & 2.9 & 6.8 & 48.9 \\
 0.5 & 0.5 & 3.2 & 9.6 & 56.3 \\
 1 & 1 & 3.4 & 7.2 & 40.6 \\
 1 & 0.5 & 2.8 & 2.6 & 46.8
\end{tabular}
\end{table}

Models do best when their training SPP matches the eval data's SPP, as expected. However, the effect of mismatch for near SI-SDRi is relatively minor. For far SI-SDRi, we observe more degradation with mismatched SPPs, indicating that perhaps source count is an important cue for far source estimation.

\subsection{Effect of silent targets}

To study the effect of silent targets, we pre-filtered our RIRs to achieve certain percentages of silent targets (before applying SPP).
Results are shown
in Table~\ref{tab:silent_targets_results} for SPPs of 1.0 and 0.5. Note that SPP and percentage of rooms with silent targets are compounding effects. As expected, showing the model more examples with non-silent targets increases SI-SDRi on non-silent examples, but degrades noise reduction on silent near examples.

\begin{table}[th]
  \caption{Performance as a function of percentage silent training targets. 
  ``\% rooms silent target'' is how many rooms in the training set had silent near or far targets,
  before the SPP was applied. Noise reduction is calculated only for the 271 eval examples with silent near targets.}
  \label{tab:silent_targets_results}
  \centering
\begin{tabular}{cc|cc|c|}
\multirow{2}{*}{\textbf{SPP}} & \multirow{2}{*}{\textbf{\begin{tabular}[c]{@{}c@{}}\% rooms\\ silent target\end{tabular}}} & \multicolumn{2}{c|}{\textbf{SI-SDRi}} & \multirow{2}{*}{\textbf{\begin{tabular}[c]{@{}c@{}}Noise \\ reduction\end{tabular}}} \\
 & & \textbf{Near} & \textbf{Far} & \\ \hline
1.0 & 0 & 6.7 & 7.8 & 23.7 \\
1.0 & 30 & 5.0 & 6.9 & 40.6 \\
0.5 & 0 & 4.6 & 6.5 & 35.4 \\
0.5 & 30 & 4.4 & 6.8 & 48.9
\end{tabular}
\end{table}

\section{Conclusions}
We proposed the novel task of separating mixtures of signals based on their distance from a single microphone.  To address this task we applied an existing high-performing sound separation model architecture, trained on synthetic data prepared to exemplify the primary qualities of sound that vary with distance.  Experiments showed promising initial performance on this task, with our model benchmark 1.5 meter distance threshold model producing an average improvement of 4.4 dB in SI-SDR for nearby sounds.  

In future work, we plan to handle more realistic scenarios with both speech and non-speech and to investigate multi-channel approaches. We also aim to elucidate the relative importance of acoustic cues, such as relative loudness and DRR, in distance-based separation networks, and evaluate their performance
in real acoustic conditions.  Training with more realistic acoustic effects, including air absorption and proximity effects, may also be useful for generalization to real world conditions, and we plan to investigate this using improved simulations.

\section{Acknowledgements} We would like to acknowledge Dick Lyon and Malcolm Slaney for helpful discussions during the preparation of this work, and Hakan Erdogan for useful comments on the manuscript.

\newpage
\bibliographystyle{IEEEbib}
\balance
\bibliography{refs}

\end{document}